\documentclass[twocolumn,floatfix,eqsecnum,prb]{revtex4}
\usepackage{graphicx}
\usepackage{amsmath}
\usepackage{amssymb}
\usepackage{bm}

\begin{document}

\title{Quantum phase transitions of a disordered antiferromagnetic topological insulator}
\author{P. Baireuther}
\affiliation{Instituut-Lorentz, Universiteit Leiden, P.O. Box 9506, 2300 RA Leiden, The Netherlands}
\author{J. M. Edge}
\affiliation{Instituut-Lorentz, Universiteit Leiden, P.O. Box 9506, 2300 RA Leiden, The Netherlands}
\author{I. C. Fulga}
\affiliation{Instituut-Lorentz, Universiteit Leiden, P.O. Box 9506, 2300 RA Leiden, The Netherlands}
\author{C. W. J. Beenakker}
\affiliation{Instituut-Lorentz, Universiteit Leiden, P.O. Box 9506, 2300 RA Leiden, The Netherlands}
\author{J. Tworzyd{\l}o}
\affiliation{Institute of Theoretical Physics, Faculty of Physics, University of Warsaw, Ho\.{z}a 69, 00--681 Warsaw, Poland}
\date{September 2013}
\begin{abstract}
We study the effect of electrostatic disorder on the conductivity of a three-dimensional antiferromagnetic insulator (a stack of quantum anomalous Hall layers with staggered magnetization). The phase diagram contains regions where the increase of disorder first causes the appearance of surface conduction (via a topological phase transition), followed by the appearance of bulk conduction (via a metal-insulator transition). The conducting surface states are stabilized by an effective time-reversal symmetry that is broken locally by the disorder but restored on long length scales. A simple self-consistent Born approximation reliably locates the boundaries of this socalled ``statistical'' topological phase.
\end{abstract}
\maketitle

\section{Introduction}
\label{intro}

Topological insulators (TI) have an insulating bulk and a conducting surface, protected by time-reversal symmetry.\cite{Has10,Xi11} In three-dimensional (3D) lattices the concept can be extended to include magnetic order:\cite{Mon10,Ess12,Fan13,Liu13,Liu13b} Antiferromagnetic topological insulators (AFTI) break time-reversal symmetry locally, but recover it in combination with a lattice translation. Layered structures with a staggered magnetization provide the simplest example of an AFTI:\cite{Mon10} The quantum anomalous Hall effect in a single layer produces edge states with a chirality that changes from one layer to the next. Interlayer coupling gives these counterpropagating edge states an anisotropic dispersion, similar to the unpaired Dirac cone on the surface of a time-reversally invariant TI --- but now appearing only on surfaces perpendicular to the layers.

While the first AFTI awaits experimental discovery, it is clear that disorder will play a essential role in any realistic material. Electrostatic disorder breaks translational symmetry, and therefore indirectly breaks the effective time-reversal symmetry of the AFTI. The topological protection of the conducting surface is expected to persist, at least for a range of disorder strengths, because the symmetry is restored on long length scales. A disordered AFTI belongs to the class of statistical topological insulators, protected by a symmetry that holds on average.\cite{Fu12,Ful12}

Here we explore these unusual disorder effects both analytically and numerically, for a simple model of a layered AFTI. We find that, while sufficiently strong disorder suppresses both bulk and surface conduction, intermediate disorder strengths may actually favor conductivity. Over a broad range of magnetizations the electrostatic disorder drives the insulating bulk into a metallic phase, via an Anderson metal-insulator transition. Disorder may also produce a topological phase transition, enabling surface conduction while keeping the bulk insulating --- as a magnetic analogue of the ``topological Anderson insulator''.\cite{Li09,Gro09,Guo10,Kob13} Each of these quantum phase transitions is identified via the scaling of the conductance with system size.

The outline of the paper is as follows. In the next section we construct a simple model of an antiferromagnetically ordered stack, starting from the Qi-Wu-Zhang Hamiltonian\cite{Qi06} for the quantum anomalous Hall effect in a single layer, and alternating the sign of the magnetization from one layer to the next. We identify the effective time-reversal symmetry of Mong, Essin, and Moore,\cite{Mon10} locate the 2D Dirac cones of surface states and the 3D Weyl cones of bulk states, and calculate their contributions to the electrical conductance. All of this is for a clean system. Disorder is added in Secs.\ \ref{phasediagram} and \ref{scaling}, where we study the quantum phase transitions between the AFTI phase and the metallic or topologically trivial insulating phases. The phase boundaries are calculated analytically using the self-consistent Born approximation, following the approach of Ref.\ \onlinecite{Gro09}, and numerically from the scaling of the conductance with system size in a tight-binding discretization of the AFTI Hamiltonian. We conclude in Sec.\ \ref{conclude}.

\section{Clean limit}
\label{cleanlimit}

\subsection{Model Hamiltonian}
\label{modelH}

\begin{figure}[tb]
\centerline{\includegraphics[width=0.8\linewidth]{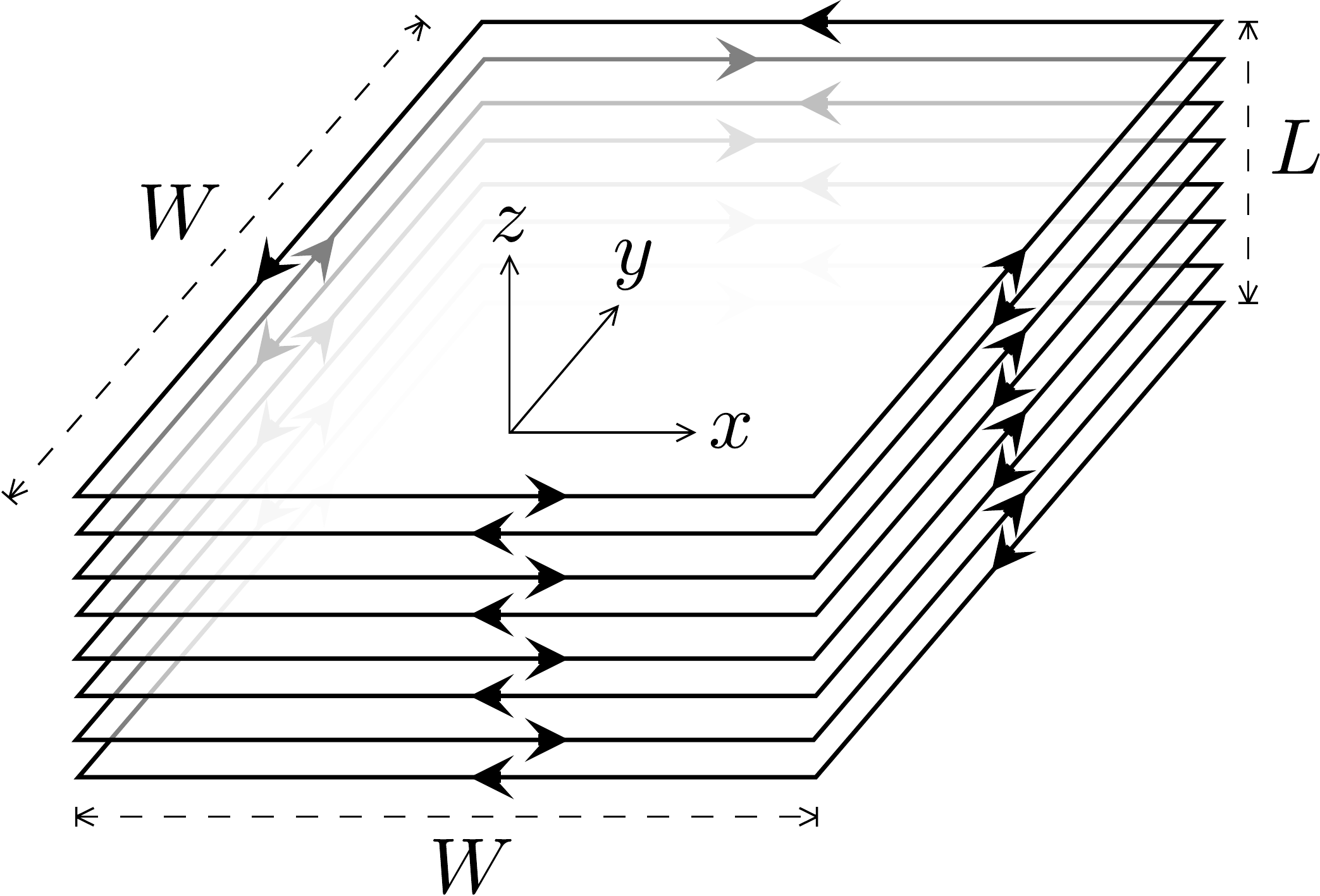}}
\caption{Stack of antiferromagnetically ordered layers. Each layer is insulating in the interior but supports a chiral edge state (arrows) because of the quantum anomalous Hall effect. Interlayer hopping (in the $z$-direction) produces an anisotropic Dirac cone of surface states on surfaces perpendicular to the layers. The unpaired Dirac cone is robust against disorder, as in a (strong) topological insulator, although time-reversal symmetry is broken locally.
}
\label{fig_layout}
\end{figure}

There exists a broad class of 3D magnetic textures that produce an AFTI.\cite{Mon10,Fan13,Liu13} Here we consider a particularly simple example of antiferromagnetically ordered layers, see Fig.\ \ref{fig_layout}, but we expect the generic features of the phase diagram to be representative of the entire class of AFTI. 

For a single layer we take the Qi-Wu-Zhang Hamiltonian of the quantum anomalous Hall effect,\cite{Qi06}
\begin{align}
H_{\pm}(k_x,k_y)={}&\pm\sigma_z(\mu-\cos k_x-\cos k_y)\nonumber\\
&+\sigma_x\sin k_x+\sigma_y\sin k_y.\label{HQWZ}
\end{align}
This is the continuum limit of a tight-binding Hamiltonian on a square lattice in the $x$-$y$ plane, with two spin bands (Pauli matrices $\bm{\sigma}$, unit matrix $\sigma_0$) coupled to the wave vector $\bm{k}$. The lattice constant and the nearest-neighbor hopping energies are set equal to unity, so that both the wave vector $\bm{k}$ and the magnetization $\mu$ are dimensionless. Time-reversal symmetry maps $H_{+}$ onto $H_{-}$,
\begin{equation}
\sigma_{y}H_{\pm}^{\ast}(-\bm{k})\sigma_{y}=H_{\mp}(\bm{k}).\label{Hpmsymmetry}
\end{equation}

The topological quantum number (Chern number) ${\cal C}_{\pm}$ of the quantum anomalous Hall Hamiltonian $H_{\pm}$ is\cite{Qi06}
\begin{equation}
{\cal C}_{\pm}=\begin{cases}
\pm\,{\rm sign}\,\mu&{\rm if}\;\;|\mu|<2,\\
0&{\rm if}\;\;|\mu|>2.
\end{cases}\label{CQWZ}
\end{equation}
A change in ${\cal C}_{\pm}$ is accompanied by a closing of the excitation gap at $\mu=-2,0,2$.

The quantum anomalous Hall layers can be stacked in the $z$-direction with ferromagnetic order (same Chern number in each layer, see Ref.\ \onlinecite{Del12}) or with antiferromagnetic order (opposite Chern number in adjacent layers). Ferromagnetic order breaks time-reversal symmetry globally, producing a 3D analogue of the quantum Hall effect with chiral surface states.\cite{Cha12,Bal12} To obtain an effective time-reversal symmetry and produce a surface Dirac cone we take an antiferromagnetic  magnetization. 

The Hamiltonian is constructed as follows. Because of the staggered magnetization, the unit cell extends over two adjacent layers, distinguished by a pseudospin degree of freedom $\tau$. The corresponding Brillouin zone is $|k_x|<\pi$, $|k_y|<\pi$, $|k_z|<\pi/2$, half as small in the $z$-direction because of the doubled unit cell. Interlayer coupling by nearest-neigbor hopping (with strength $t_{z}$) is described by the Hamiltonian
\begin{equation}
H_{z}(k_z)=t_{z}\begin{pmatrix}
0&\rho^{\dagger}e^{2ik_z}+\rho\\
\rho e^{-2ik_z}+\rho^\dagger&0
\end{pmatrix},\label{Hinter}
\end{equation}
with a $2\times 2$ matrix $\rho$ acting on the spin degree of freedom. The term $\rho^{\dagger}e^{2ik_z}$ moves up one layer in the next unit cell, while the term $\rho$ moves down one layer in the same unit cell. We require that the interlayer Hamiltonian preserves time-reversal symmetry,
\begin{equation}
\sigma_{y}H^{\ast}_{z}(-k_{z})\sigma_{y}=H_{z}(k_{z})\Rightarrow\sigma_{y}\rho^{\ast}\sigma_{y}=\rho.\label{tildeHz}
\end{equation}
This still leaves some freedom in the choice of $\rho$, we take $\rho=i\sigma_{z}$.

The staggered magnetization is described by combining $H_{+}$ in one layer with $H_{-}$ in the next layer, so by replacing $\sigma_z$ with $\tau_{z}\otimes\sigma_{z}$ in Eq.\ \eqref{HQWZ}. [The Pauli matrices $\bm{\tau}$ (unit matrix $\tau_0$) act on the layer degree of freedom.] The full Hamiltonian of the stack takes the form
\begin{align}
&H_{\rm AFTI}(\bm{k})=H_{z}(k_z)+(\tau_z\otimes\sigma_z)(\mu-\cos k_x-\cos k_y)\nonumber\\
&\quad\quad\quad+\tau_0\otimes(\sigma_x\sin k_x+\sigma_y\sin k_y),\label{HAFTI}\\
&H_{z}(k_{z})=t_{z}(\tau_y\otimes\sigma_{z})(\cos 2k_z-1)+t_{z}(\tau_{x}\otimes\sigma_{z})\sin 2k_z.\label{Hzspecialchoice}
\end{align}

\subsection{Effective time-reversal symmetry}
\label{effectiveTRS}

Following Mong, Essin, and Moore,\cite{Mon10} we construct an effective time-reversal symmetry operator,
\begin{equation}
{\cal S}(k_z)=\Theta {\cal T}(k_z)= {\cal T}(k_z)\Theta,\label{calSdef}
\end{equation}
by combining the fundamental time-reversal operation $\Theta$ with a translation ${\cal T}(k_z)$ over half a unit cell in the $z$-direction. The translation operator is represented by a $2\times 2$ matrix acting on the layer degree of freedom, 
\begin{equation}
{\cal T}(k_z)=\begin{pmatrix}
0&e^{2ik_z}\\
1&0
\end{pmatrix}=e^{ik_z}(\tau_x\cos k_z-\tau_y\sin k_z).
\end{equation}
Both off-diagonal matrix elements switch the layers, either remaining in the same unit cell or moving to the next unit cell. One verifies that the square ${\cal T}^{2}(k_z)=e^{2ik_z}\tau_0$ represents the Bloch phase acquired by a shift over the full unit cell (two layers).

The interlayer Hamiltonian \eqref{Hinter} commutes with the translation over half a unit cell,
\begin{equation}
{\cal T}(k_z)H_{z}(k_z)=H_z(k_z){\cal T}(k_z).\label{THcommute}
\end{equation}
Since we have also assumed that $H_z$ preserves time-reversal symmetry, $\Theta H_z(k_z)=H_z(k_z)\Theta$, it commutes with the combined operation,
\begin{equation}
{\cal S}(k_z)H_z(k_z)=H_z(k_z){\cal S}(k_z).\label{SHcommute}
\end{equation}
The full Hamiltonian,
\begin{equation}
H_{\rm AFTI}(\bm{k})=H_{z}(k_z)+\begin{pmatrix}
H_{+}(k_x,k_y)&0\\
0&H_{-}(k_x,k_y)
\end{pmatrix},\label{Hfull}
\end{equation}
then also commutes with ${\cal S}(k_z)$, because
\begin{equation}
\Theta H_{+}(k_x,k_y)=H_{-}(k_x,k_y)\Theta.\label{ThetaHpm}
\end{equation}

For the quantum anomalous Hall layers the fundamental time-reversal operation is
\begin{equation}
\Theta = i\sigma_y {\cal K},\label{Thetadef}
\end{equation}
where ${\cal K}$ takes the complex conjugate and inverts the momenta, ${\cal K}f(\bm{k})=f^{\ast}(-\bm{k})$. [One verifies that the identity \eqref{ThetaHpm} is equivalent to Eq.\ \eqref{Hpmsymmetry}.] The effective time-reversal symmetry operation is then given explicitly by
\begin{equation}
{\cal S}(k_z)=i\sigma_{y}\otimes(\tau_x\cos k_z-\tau_y\sin k_z){\cal K},\label{Sexplicit}
\end{equation}
up to an irrelevant phase factor $e^{ik_z}$.

The fundamental time-reversal operation \eqref{Thetadef} squares to $-1$, as it should do for a spin-$\tfrac{1}{2}$ degree of freedom. As noted by Liu,\cite{Liu13} one can equally well start from a spinless time-reversal symmetry that squares to $+1$, for example, taking $\Theta={\cal K}$. Since ${\cal S}^{2}(k_z)=e^{2ik_z}\Theta^2$, the choice of $\Theta^2=\pm 1$ amounts to shift of $k_z$ by $\pi/2$. Gapless surface states appear at the $k_z$-value for which ${\cal S}$ squares to $-1$, so at the center of the surface Brillouin zone ($k_z=0$) for $\Theta^2=-1$ and at the edge ($k_z=\pi/2$) for $\Theta^2=1$.

\subsection{Bulk and surface states}
\label{bulk_surface_states}

The bulk spectrum $E(\bm{k})$ of the Hamiltonian \eqref{HAFTI} can be easily calculated by noting that $H_{\rm AFTI}^2(\bm{k})$ reduces to a unit matrix in $\sigma,\tau$ space, hence
\begin{align}
E^{2}(\bm{k})={}&(\mu-\cos k_x-\cos k_y)^2+\sin^{2}k_x+\sin^{2}k_y\nonumber\\
&+(2t_z\sin k_z)^2.\label{bulkspectrum}
\end{align}
The gap closes with a 3D conical dispersion (Weyl cone) at $(k_x,k_y,k_z)=(0,0,0)$ for $\mu=2$, at $(\pi,\pi,0)$ for $\mu=-2$, and at the two points $(0,\pi,0)$, $(\pi,0,0)$ for $\mu=0$. Each cone is twofold degenerate and has the anisotropic dispersion
\begin{equation}
E_{\rm Weyl}^{2}(\delta\bm{k})=(\delta k_{x})^{2}+(\delta k_{y})^{2}+4t_{z}^{2}(\delta k_{z})^{2},\label{Econe}
\end{equation}
with $\delta \bm{k}$ the wave vector measured from the conical point (Weyl point). Unlike in the case of ferromagnetic order,\cite{Del12,Bur11} the bulk spectrum is only gapless at specific values of $\mu\in\{0,\pm 2\}$ --- there is no Weyl semimetal phase in this model. 

The surface spectrum of the antiferromagnetically ordered stack is gapless in the interval $0<|\mu|<2$, if finite-size effects are avoided by taking periodic boundary conditions in the $z$-direction. The surface states have an anisotropic 2D conical dispersion (Dirac cone),
\begin{equation}
\begin{split}
E_{\rm Dirac}^{2}(q,k_z)&=(q-q_0)^{2}+4t_{z}^{2}k_{z}^{2},\\
&q_0=\begin{cases}
0&{\rm if}\;\;0<\mu<2,\\
\pi&{\rm if}\;\;-2<\mu<0,
\end{cases}
\end{split}\label{Dirac}
\end{equation}
with $q=k_x$ on the $x$-$z$ plane and $q=k_y$ on the $y$-$z$ plane.

These AFTI surface states emerge from the counterpropagating chiral edge states at $k_z=0$ and are protected by the effective time-reversal symmetry \eqref{Sexplicit}. They are reminiscent of the surface states in a weak topological insulator, formed by stacking quantum spin Hall layers with helical edge states. The essential difference is that in a weak TI there is a second Dirac cone at $k_z=\pi$, while the AFTI has only a single Dirac cone. (The ``fermion doubling'' is avoided by the restriction of the Brillouin zone to $|k_z|<\pi/2$.)

\begin{figure}[tb]
\centerline{\includegraphics[width=1\linewidth]{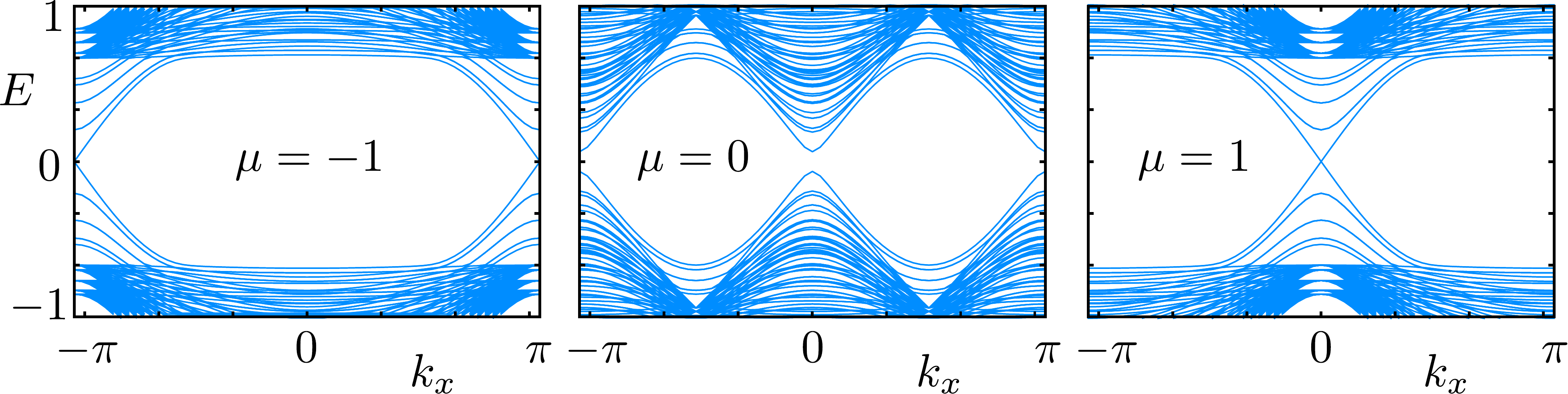}}
\caption{Energy spectrum of the AFTI Hamiltonian \eqref{HAFTI}, with $t_{z}=0.4$, for a stack of 16 layers in the $z$-direction with periodic boundary conditions. The layers are infinitely wide in the $x$-direction and truncated at 16 lattice sites in the $y$-direction. At $\mu=\pm 1$ the system is in the AFTI phase, with a nondegenerate Dirac cone of surface states centered at the edge of the Brillouin zone ($-2<\mu<0$) or at the center of the Brillouin zone ($0<\mu<2$). At $\mu=0$ the bulk gap closes at a pair of twofold degenerate Weyl cones, one at the center and one at the edge of the Brillouin zone. In this plot a finite gap remains for $\mu=0$, because of the confinement in the $y$-direction.
}
\label{fig_cones}
\end{figure}

Notice that the closing of the gap at $\mu=0$ is not accompanied by a change in the number of surface Dirac cones. Instead, the single Dirac cone switches from the center to the edge of the surface Brillouin zone when $\mu$ crosses zero. (See Fig.\ \ref{fig_cones}.) This is a quantum phase transition in the sense of Ref.\ \onlinecite{Sla13}, between band insulators with the same topological quantum number but distinguished by the location of the surface Dirac cone. 

\subsection{Surface conductance from the Dirac cone}
\label{Dirac_conductance}

To study the transport properties of the AFTI, we take layers in the $x$-$y$ plane of width $W\times W$, stacked in the $z$-direction over a length $L$. The top and bottom layers are connected to electron reservoirs at voltage difference $V$, and the current $I$ in the $z$-direction then determines the conductance $G=\lim_{V\rightarrow 0}I/V$ perpendicular to the layers. We fix the Fermi level $E_F=0$ at the middle of the bulk gap, where the conductance is minimal.

In the AFTI phase, for $0<|\mu|<2$, the conductance is dominated by the surface states. Analogously to graphene,\cite{Bee08,Kat12} each 2D Dirac cone contributes a conductance $(e^2/\pi h)(W/L_{\rm eff})$, at the Dirac point ($E_F=0$) and for $W\gg L_{\rm eff}\equiv L/2t_z$. There are four Dirac cones (one on each surface perpendicular to the layers), totaling
\begin{equation}
G_{\rm Dirac}=\frac{8e^{2}}{\pi h}\frac{t_{z}W}{L}.\label{GDirac}
\end{equation}

\subsection{Bulk conductance from the Weyl cone}
\label{Weyl_conductance}

When the bulk gap closes, at $\mu=0,\pm 2$, the 3D Weyl cones contribute an amount of order $(W/L_{\rm eff})^2$ to the conductance, which dominates over the surface conductance when $W\gg L_{\rm eff}$. A similar calculation as in Ref.\ \onlinecite{Two06} gives the minimal conductance at the Weyl point ($E_F=0$),
\begin{align}
&G_{\rm Weyl}=d\,\frac{e^{2}}{h}\sum_{n,m=-\infty}^{\infty}T_{nm},\label{GWeyl}\\
&T_{nm}=\cosh^{-2}\left[2\pi(L_{\rm eff}/W)\sqrt{n^{2}+m^{2}}\right],\label{Tnmdef}
\end{align}
for periodic boundary conditions in the $x$ and $y$-directions. Four Weyl cones contribute at $\mu=0$ (degeneracy factor $d=4$) and two Weyl cones contribute at $\mu=\pm 2$ (degeneracy factor $d=2$).

\begin{figure}[tb]
\centerline{\includegraphics[width=0.8\linewidth]{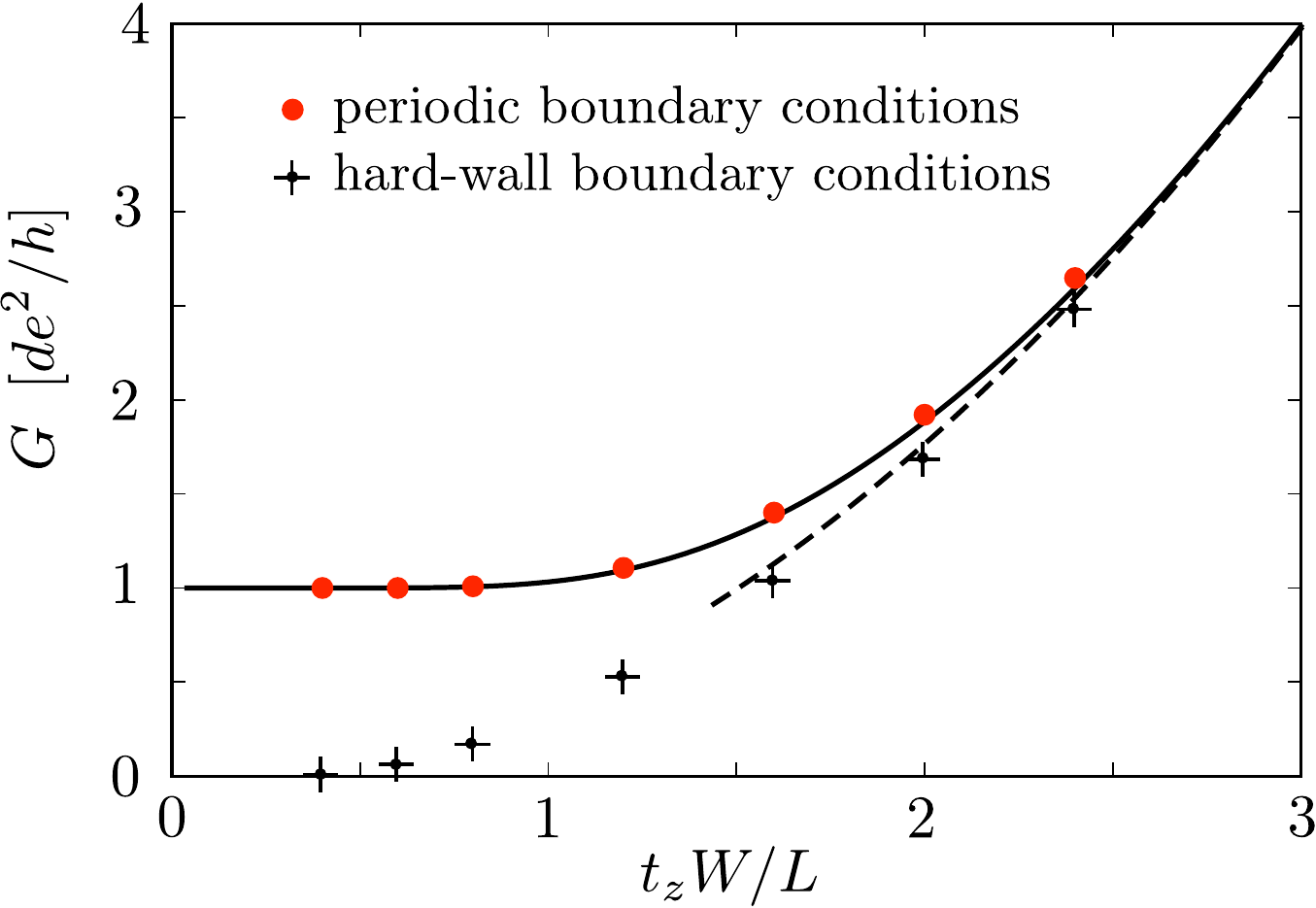}}
\caption{Conductance at the Weyl point for periodic boundary conditions, according to Eq.\ \eqref{GWeyl} (solid curve) and the asymptotic form for large aspect ratio \eqref{GWeylsimple} (dashed). The data points are calculated from the AFTI Hamiltonian \eqref{HAFTI}, at $\mu=2$, $t_{z}=0.4$, for a lattice of 8 layers in the $z$-direction, with periodic boundary conditions in the $x$ and $y$-directions (red dots) and for hard-wall boundary conditions (black crosses). 
}
\label{fig_Weyl}
\end{figure}

The dependence of $G_{\rm Weyl}$ on the aspect ratio $W/L_{\rm eff}$ is plotted in Fig.\ \ref{fig_Weyl}. For $W\gg L_{\rm eff}$ one has the asymptotic result
\begin{equation}
G_{\rm Weyl}=d\frac{e^{2}}{h}\frac{2\ln 2}{\pi}\left(\frac{t_z W}{L}\right)^{2}.\label{GWeylsimple}
\end{equation}

\begin{figure}[tb]
\centerline{\includegraphics[width=0.8\linewidth]{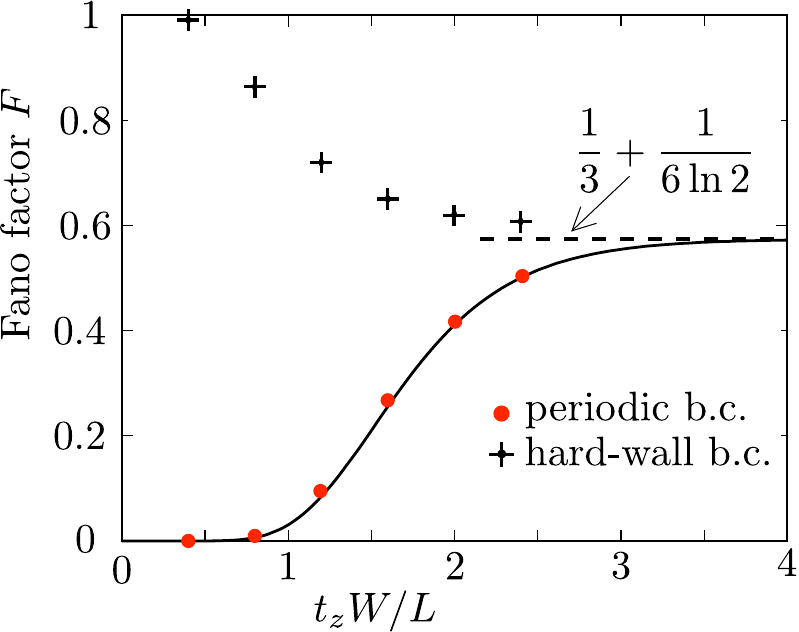}}
\caption{Same as Fig.\ \ref{fig_Weyl}, but for the Fano factor at the Weyl point.
}
\label{fig_WeylFano}
\end{figure}

The conduction at the Weyl point is not ``pseudo-diffusive'', as it is at the Dirac point of graphene, because the conductivity $\sigma_{\rm Weyl}=G_{\rm Weyl}L/W^2$ is not scale invariant. The Fano factor $F_{\rm Weyl}$ (ratio of shot noise power and average current) at the Weyl point is scale invariant, but it differs from the value $F=1/3$ characteristic of pseudo-diffusive conduction.\cite{Two06} We find
\begin{align}
F_{\rm Weyl}&=\frac{\sum_{n,m=-\infty}^{\infty}T_{nm}(1-T_{nm})}{\sum_{n,m=-\infty}^{\infty}T_{nm}}\nonumber\\
&=\frac{1}{3}+(6\ln 2)^{-1}\approx 0.574\;\;{\rm for}\;\;W\gg L_{\rm eff}.\label{FWeyl}
\end{align}
The aspect ratio dependence of $F_{\rm Weyl}$ is plotted in Fig.\ \ref{fig_WeylFano}.

\section{Phase diagram of the disordered system}
\label{phasediagram}

We add disorder to the AFTI Hamiltonian \eqref{HAFTI} in the form of a spin-independent random potential chosen independently on each lattice site from a Gaussian distribution of zero mean and variance $\delta U^2$. In $\sigma,\tau$ representation the disorder Hamiltonian is given by
\begin{align}
&H_{\rm disorder}=\sum_{i}\left[(\tau_{0}\otimes\sigma_{0})U_{i}^{(1)}+(\tau_{z}\otimes\sigma_{0})U_{i}^{(2)}\right],\label{Hdisordersigmatau}\\
&\langle U_{i}^{(n)}\rangle=0,\;\;\langle U_{i}^{(n)}U_{i'}^{(n')}\rangle=\tfrac{1}{2}\delta U^{2}\delta_{ii'}\delta_{nn'}.\label{Ucorrelator}
\end{align}
The sum over $i$ runs over bilayer unit cells and $\langle\cdots\rangle$ denotes the disorder average. 

Different layers see a different random potential, so the effective time-reversal symmetry of Sec.\ \ref{effectiveTRS} is broken locally by the disorder --- but restored on long length scales. We expect the effect on the AFTI of a random potential to be equivalent to the effect on a strong TI of a random magnetic field:\cite{Ful12,Nom08} The surface remains conducting while the bulk remains insulating, separated from the trivial insulator by a topological phase transition.

In this section we explore the phase diagram of the disordered AFTI, first analytically using the self-consistent Born approximation (SCBA) and then numerically by calculating the conductance.

We calculate the disorder-averaged density of states from the self-energy $\Sigma$, defined by
\begin{align}
&\frac{1}{E_{F}+i0^{+}-H_{\rm AFTI}-\Sigma}\nonumber\\
&\quad\quad=\left\langle\frac{1}{E_F+i0^{+}-H_{\rm AFTI}-H_{\rm disorder}}\right\rangle.\label{Sigmadef}
\end{align}
We set the Fermi level at $E_{F}=0$, in the middle of the gap of the clean system. The SCBA self-energy, for a disorder potential of the form \eqref{Hdisordersigmatau}, is given by the equation
\begin{align}
\Sigma={}&\tfrac{1}{2}\delta U^{2}\sum_{\bm{k}}\biggl([i0^{+}-H_{\rm AFTI}(\bm{k})-\Sigma]^{-1}\nonumber\\
&+\tau_{z}[i0^{+}-H_{\rm AFTI}(\bm{k})-\Sigma]^{-1}\tau_{z}\biggr).\label{SCBA0}
\end{align}
The sum over $\bm{k}$ ranges over the first Brillouin zone, in the continuum limit
\begin{equation}
\sum_{\bm{k}}\mapsto \frac{1}{4\pi^3}\int_{-\pi}^{\pi}dk_x\int_{-\pi}^{\pi}dk_y\int_{-\pi/2}^{\pi/2}dk_z.\label{intsum}
\end{equation}

The SCBA self-energy is a $\bm{k}$-independent $4\times 4$ matrix in the spin and layer degrees of freedom,
\begin{equation}
\Sigma=(\tau_{z}\otimes\sigma_{z})\delta\mu-(\tau_{0}\otimes\sigma_{0})i\gamma.\label{Sigmasigmatau}
\end{equation}
The term $\delta\mu$ renormalizes the magnetization $\mu$ and thus accounts for a disorder-induced shift of the phase boundary of the topologically nontrivial band insulator. The term $\gamma$ produces a density of states $\pi^{-1}{\rm Im}\,(H_{\rm AFTI}+\Sigma)^{-1}$, induced by the disorder within the gap of the clean system. A nonzero $\gamma$ may indicate a metallic phase or a topologically trivial Anderson insulator (the density of states cannot distinguish between the two).

Substitution of Eq.\ \eqref{Sigmasigmatau} into Eq.\ \eqref{SCBA0}, and use of the identity
\begin{align}
&H_{\rm AFTI}(k_x,k_y,k_z)+\tau_{z}H_{\rm AFTI}(-k_x,-k_y,k_z)\tau_{z}\nonumber\\
&\quad\quad=2(\tau_z\otimes\sigma_z)(\mu-\cos k_x-\cos k_y),\label{identity}
\end{align}
produces two coupled equations for $\gamma$ and $\delta\mu$:
\begin{subequations}
\label{SCBA}
\begin{align}
&\gamma=\delta U^{2}\sum_{\bm{k}}\frac{\gamma+0^{+}}{\gamma^2+E_{\mu+\delta\mu}^{2}(\bm{k})},\label{gamma}\\
&\delta\mu=-\delta U^{2}\sum_{\bm{k}}\frac{M_{\mu+\delta\mu}(\bm{k})}{\gamma^{2}+E_{\mu+\delta\mu}^{2}(\bm{k})},\label{deltamu}
\end{align}
\end{subequations}
with the definitions
\begin{subequations}
\label{EMmu}
\begin{align}
&E_{\mu}^{2}(\bm{k})=M_{\mu}^{2}(\bm{k})+\sin^{2}k_x+\sin^{2}k_y+4t^{2}_z\sin^{2} k_z,\label{Emu}\\
&M_{\mu}(\bm{k})=\mu-\cos k_x-\cos k_y.\label{Mmu}
\end{align}
\end{subequations}

\begin{figure}[tb]
\centerline{\includegraphics[width=1\linewidth]{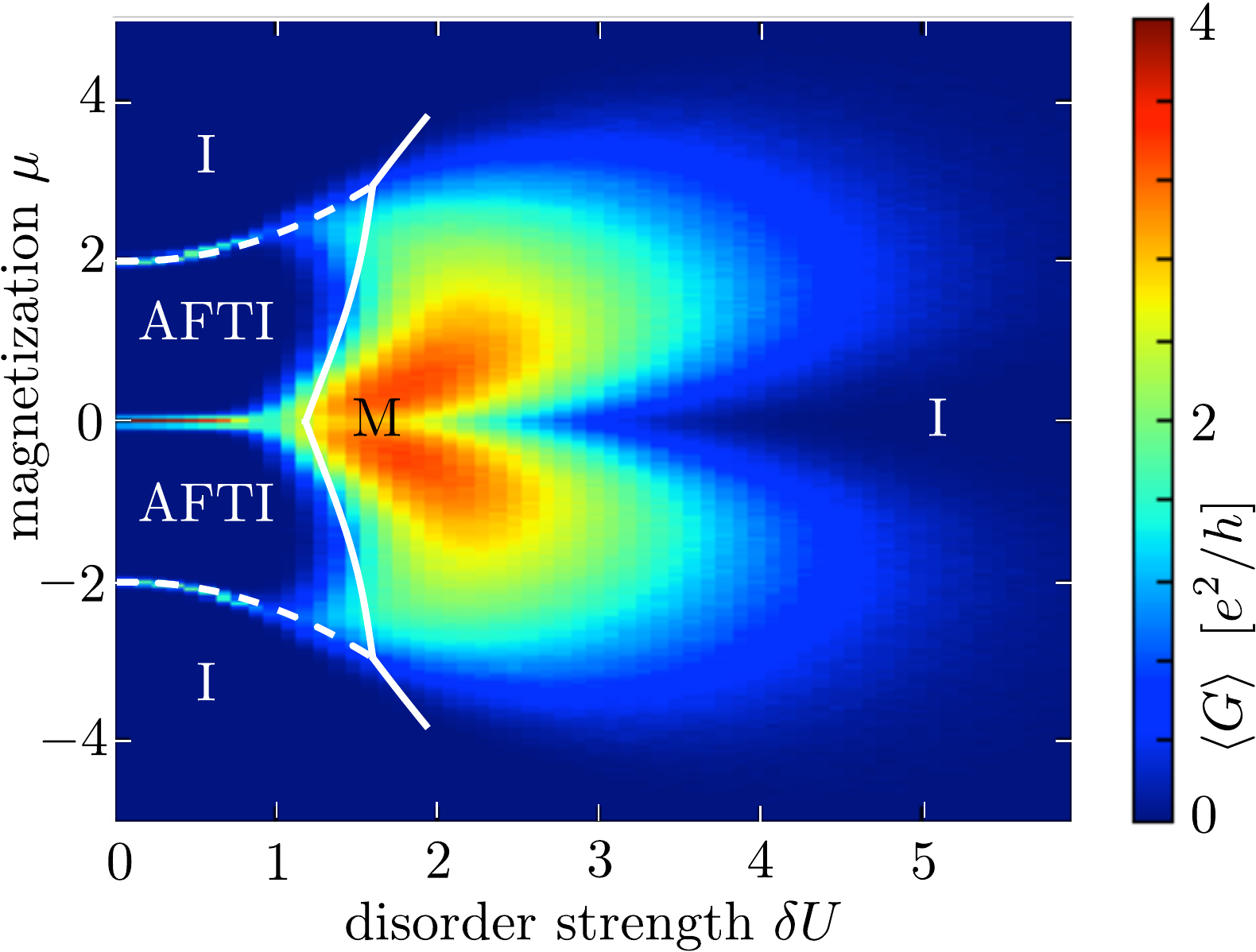}}
\caption{Color-scale plot of the conductance of a disordered AFTI, calculated numerically from the Hamiltonian \eqref{HAFTI} for current flowing perpendicular to a stack of 20 layers. Each layer has dimensions $20\times 20$ with periodic boundary conditions, the interlayer coupling is $t_{z}=0.4$. The topological insulator phase (AFTI), the trivial insulator phase (I), and the metallic phase (M) are indicated in the plot. The white curves are the phase boundaries resulting from the self-consistent Born approximation (SCBA). The Anderson transition between a metal and a trivial insulator is not captured by the SCBA.
}
\label{fig_phasediagram}
\end{figure}

The phase boundary at $\mu=0$ remains unaffected by disorder, because 
\begin{equation}
\sum_{\bm{k}}\frac{M_{0}(\bm{k})}{E_{0}^{2}(\bm{k})}=0,\label{summuis0}
\end{equation}
so $\gamma=0=\delta\mu$ solves the SCBA equations for $\mu=0$. The phase boundaries at $\mu=\pm 2$ do shift when we switch on the disorder. If we seek a solution of Eq.\ \eqref{SCBA} with $\gamma=0$, $\delta\mu=\pm 2-\mu_{\pm}$ we obtain the phase boundaries at
\begin{equation}
\mu_{\pm}=\pm 2+\delta U^{2}\sum_{\bm k}\frac{M_{\pm 2}(\bm{k})}{E_{\pm 2}^{2}(\bm{k})}.\label{muphaseboundary}
\end{equation}
These phase boundaries between band insulators are plotted in Fig.\ \ref{fig_phasediagram} (dashed curves), at the value $t_{z}=0.4$ for which $\mu_{\pm}=\pm 2\pm 0.345\,\delta U^{2}$.

The outward curvature of the phase boundaries implies that the addition of disorder to a topologically trivial insulator can convert it into a nontrivial insulator, or in other words, that disorder can produce metallic conduction on surfaces perpendicular to the layers --- analogous to a topological Anderson insulator.\cite{Li09,Gro09,Guo10,Kob13}

For sufficiently large $\delta U>\delta U_{c}$, the SCBA equations may support a solution with nonzero $\gamma$. The dependence of $\delta U_{c}$ on $\mu$ follows from the solution of Eq.\ \eqref{SCBA} for infinitesimal $\gamma\neq 0$,
\begin{equation}
\delta U_{c}^{2}=\left[\sum_{\bm{k}}\frac{1}{E_{x}^{2}(\bm{k})}\right]^{-1},\;\;\mu=x+\delta U_{c}^{2}\sum_{\bm{k}}\frac{M_{x}(\bm{k})}{E_{x}^{2}(\bm{k})}.\label{muc}
\end{equation}
By varying $x\equiv\mu+\delta\mu$ we obtain the phase boundary $\delta U_{c}(\mu)$ plotted in Fig.\ \ref{fig_phasediagram} (solid curve), separating the band insulator from a metallic phase (or possibly an Anderson insulator with a finite density of states in the band gap). 

At $x=\pm 2$ we reach a tricritical point, where the metal meets two topologically distinct insulating phases. For $t_{z}=0.4$ these tricritical points occur at $\mu=\pm 2.940$, $\delta U_{c}=1.654$.

We have tested the SCBA by calculating the conductance from the AFTI Hamiltonian \eqref{HAFTI}, discretized on a cubic lattice of dimensions $W\times W\times L=20\times 20\times 20$. (These numerical calculations were performed using the Kwant code.\cite{kwant}) We impose periodic boundary conditions in the $x$ and $y$-directions and connect the layers at $z=0$ and $z=L$ to $W^2$ one-dimensional chains, as a model of a heavily doped electron reservoir. The interlayer coupling is fixed at $t_z=0.4$. The conductance, averaged over a few hundred disorder realizations, is shown as a color-scale plot in Fig.\ \ref{fig_phasediagram}.

As expected, the SCBA cannot describe the phase boundary between the trivial insulator and the metal, since it cannot distinguish between insulating and extended states in the bulk gap. For the other phase boundaries, between the topologically trivial and nontrivial insulators (dashed) as well as between the nontrivial insulator and the metal (solid), the SCBA is found to be in good agreement with the conductance calculations.

\section{Finite-size scaling}
\label{scaling}

\begin{figure}[tb]
\centerline{\includegraphics[width=1\linewidth]{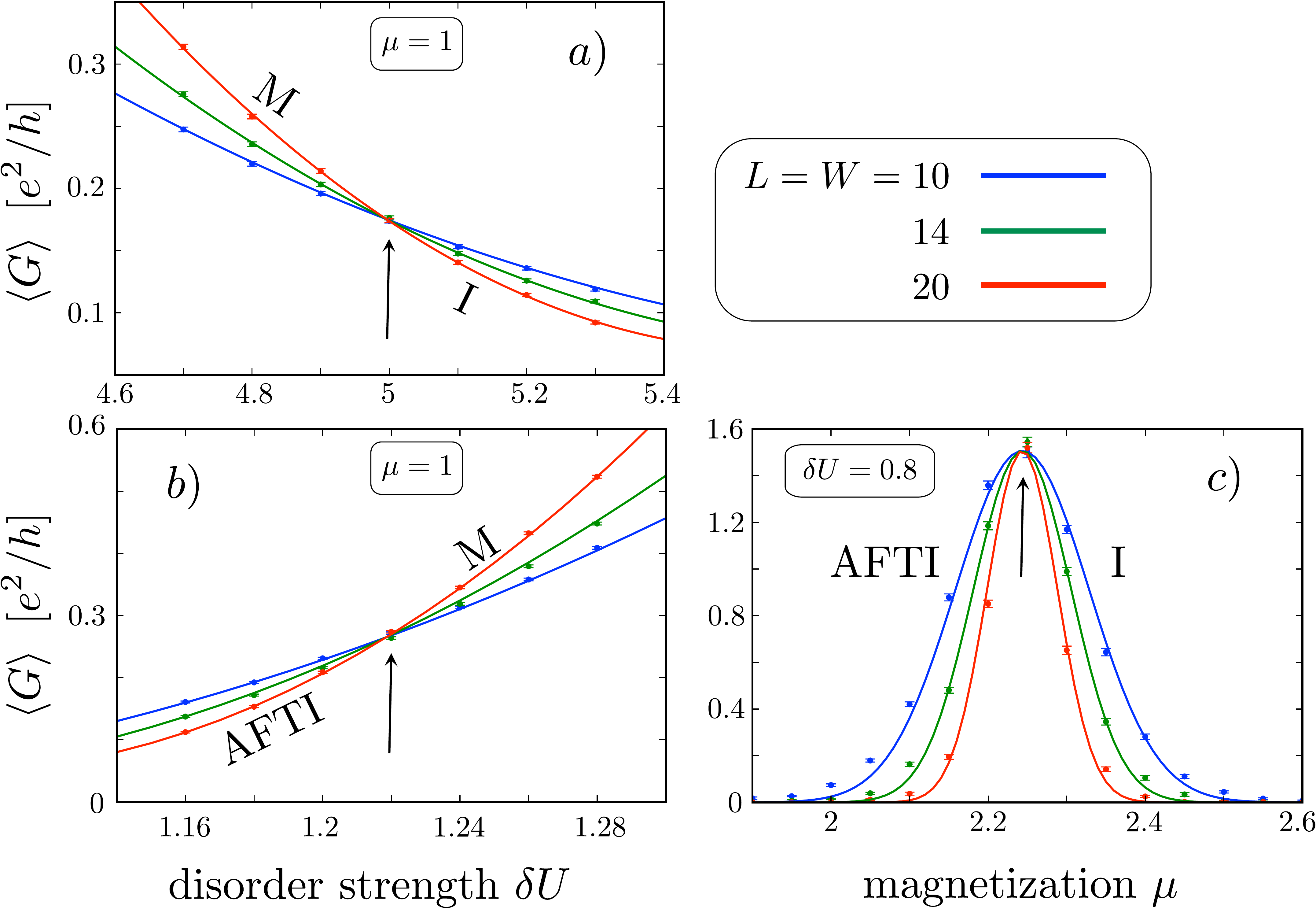}}
\caption{Disorder averaged conductance for three system sizes. Panels \textit{a} and \textit{b} show the transition between a metal (M) and an insulator which is topologically trivial (I) or nontrivial (AFTI). Panel \textit{c} show the trivial-to-nontrivial insulator transition. The scale-independent conductance at the critical point of the phase transition is indicated by an arrow. The curves are guides to the eye. Data points from panels $a$ and $b$ are averages over 20000 disorder configurations, data points from panel $c$ are averages over 200 configurations.}
\label{fig_scaling}
\end{figure}

The conductance in the phase diagram of Fig.\ \ref{fig_phasediagram} is given for a single size of the conductor. To establish the metallic or insulating character of a phase it is necessary to compare different system sizes. A phase transition is then identified by a scale invariant ``critical'' conductance. 

Such finite-size scaling plots are shown in Fig.\ \ref{fig_scaling}. Panel \textit{a} shows the transition from a metal to an insulator with increasing disorder, while panel \textit{b} shows the reverse transition. Panel \textit{c} shows the transition between a topologically trivial and nontrivial insulator. The critical point of each transition is indicated by an arrow.

\begin{figure}[tb]
\centerline{\includegraphics[width=0.8\linewidth]{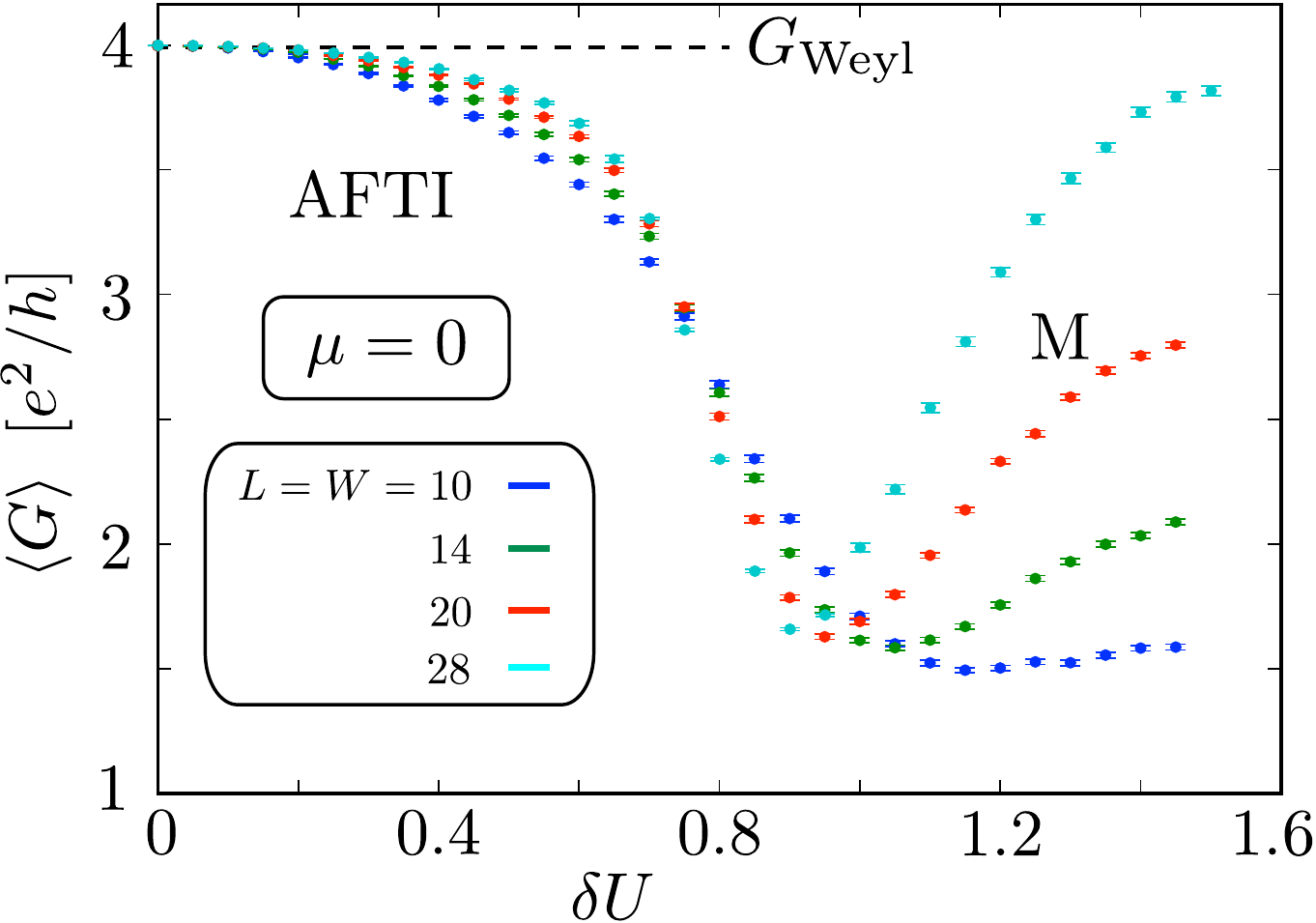}}
\caption{Disorder averaged conductance on the line $\mu=0$, within the AFTI phase for weak disorder and metallic for strong disorder. The ballistic conductance at the Weyl point is indicated.
}
\label{fig_Weylscaling}
\end{figure}

The finite-size scaling on the line $\mu=0$ is shown in Fig.\ \ref{fig_Weylscaling}. For weak disorder the conductance tends to saturate with increasing system size at the clean limit \eqref{GWeyl}, which for $d=4$, $t_{z}=0.4$, and $W=L$ is close to $G_{\rm Weyl}=4e^{2}/h$. For strong disorder the conductance shows the metallic scaling $\propto W^2/L=L$, but only after an intermediate regime where the conductance decreases with increasing system size --- suggestive of an insulating regime. We will discuss the implications in the next section.

\section{Discussion}
\label{conclude}

We have investigated how disorder affects the phase diagram of a simple model in the class of antiferromagnetic topological insulators.\cite{Mon10} Depending on the disorder strength, topologically trivial (I) or nontrivial (AFTI) phases appear, as well as a metallic phase (M). The I-AFTI and M-AFTI phase boundaries are well described by the self-consistent Born approximation (dashed and solid curves in Fig.\ \ref{fig_phasediagram}), including the location of the tri-critical point at which all three phases meet.

Without disorder, there is also an AFTI-AFTI transition at magnetization $\mu=0$. When the sign of $\mu$ changes, the surface Dirac cone switches from the center to the edge of the Brillouin zone (Fig.\ \ref{fig_cones}). Precisely at the transition, the bulk gap closes and a Weyl cone appears with a scale-invariant conductance $G_{\rm Weyl}$ (Fig.\ \ref{fig_Weyl}) and Fano factor $F_{\rm Weyl}$ (Fig.\ \ref{fig_WeylFano}). Since the AFTI has a $\mathbb{Z}_2$ topological quantum number, there cannot be two topologically distinct nontrivial phases. We would expect disorder to open up a pathway of localized states in the phase diagram, that would connect the AFTI phases at positive and negative magnetization.

The numerical calculations in Fig.\ \ref{fig_Weylscaling} show an indication of this localized regime on the line $\mu=0$, for disorder strengths around $\delta U\approx 0.8$, before the transition into a metallic phase at stronger disorder. The limited range of system sizes does not allow for a conclusive identification, but the numerics is consistent with our expectation of one single topologically nontrivial phase.
 
In conclusion, we have demonstrated that the notion of an antiferromagnetic topological insulator,\cite{Mon10} protected by the effective $k$-dependent time-reversal symmetry \eqref{calSdef}, extends to disordered systems where momentum $k$ is no longer a good quantum number. The system then belongs to the class of \textit{statistical} topological insulators,\cite{Fu12,Ful12} protected by an ensemble-averaged symmetry.
 
\acknowledgments

Discussions with A. R. Akhmerov are gratefully acknowledged. This work was supported by the Dutch Science Foundation NWO/FOM, by an ERC Advanced Investigator Grant, and by the EU Network NanoCTM.


\begin{thebibliography}{99}
\bibitem{Has10} M. Z. Hasan and C. L. Kane, Rev. Mod. Phys. \textbf{82}, 3045 (2010).
\bibitem{Xi11} X.-L. Qi and S.-C. Zhang, Rev. Mod. Phys. \textbf{83}, 1057 (2011).
\bibitem{Mon10} R. S. K. Mong, A. M. Essin, and J. E. Moore, Phys. Rev. B \textbf{81}, 245209 (2010).
\bibitem{Ess12} A. M. Essin and V. Gurarie, Phys. Rev. B \textbf{85}, 195116 (2012).
\bibitem{Fan13} C. Fang, M. J. Gilbert, and B. A. Bernevig, arXiv:1304.6081.
\bibitem{Liu13} C.-X. Liu, arXiv:1304.6455.
\bibitem{Liu13b} C.-X. Liu and R.-X. Zhang, arXiv:1308.4717.
\bibitem{Fu12} L. Fu and C. L. Kane, Phys. Rev. Lett. \textbf{109}, 246605 (2012).
\bibitem{Ful12} I. C. Fulga, B. van Heck, J. M. Edge, and A. R. Akhmerov, arXiv:1212.6191.
\bibitem{Li09} J. Li, R.-L. Chu, J. K. Jain, and S.-Q. Shen, Phys. Rev. Lett. \textbf{102}, 136806 (2009).
\bibitem{Gro09} C. W. Groth, M. Wimmer, A. R. Akhmerov, J. Tworzyd{\l}o, and C. W. J. Beenakker, Phys. Rev. Lett. \textbf{103}, 196805 (2009).
\bibitem{Guo10} H.-M. Guo, G. Rosenberg, G. Refael, and M. Franz, Phys. Rev. Lett. \textbf{105}, 216601 (2010).
\bibitem{Kob13} K. Kobayashi, T. Ohtsuki, and K.-I. Imura, Phys. Rev. Lett. \textbf{110}, 236803 (2013).
\bibitem{Qi06} X.-L. Qi, Y.-S. Wu, and S.-C. Zhang, Phys. Rev. B \textbf{74}, 085308 (2006).
\bibitem{Del12} P. Delplace, J. Li, and D. Carpentier, EPL \textbf{97}, 67004 (2012).
\bibitem{Cha12} J. T. Chalker and A. Dohmen, Phys. Rev. Lett. \textbf{75}, 4496 (2012).
\bibitem{Bal12} L. Balents and M. P. A. Fisher, Phys. Rev. Lett. \textbf{76}, 2782 (2012).
\bibitem{Bur11} A. A. Burkov and L. Balents, Phys. Rev. Lett. \textbf{107}, 127205 (2011).
\bibitem{Sla13} R.-J. Slager, A. Mesaros, V. Juricic, and J. Zaanen, Nature Phys. \textbf{9}, 98 (2013).
\bibitem{Bee08} C. W. J. Beenakker, Rev. Mod. Phys. \textbf{80}, 1337 (2008).
\bibitem{Kat12} M. I. Katsnelson, \textit{Graphene: Carbon in Two Dimensions} (Cambridge University Press, 2012).
\bibitem{Two06} J. Tworzyd{\l}o, B. Trauzettel, M. Titov, A. Rycerz, and C. W. J. Beenakker, Phys. Rev. Lett. \textbf{96}, 246802 (2006).
\bibitem{Nom08} K. Nomura, S. Ryu, M. Koshino, C. Mudry, and A. Furusaki, Phys. Rev. Lett. \textbf{100}, 246806 (2008).
\bibitem{kwant} C. W. Groth, M. Wimmer, A. R. Akhmerov, and X. Waintal, arXiv:1309.2926.
\end{thebibliography}
\end{document}